\documentclass[12pt,a4paper]{article}
\usepackage{amssymb}

\usepackage[left=5em]{geometry}
\usepackage{amsmath}
\usepackage{graphicx}

\title{Stability of curvature perturbation with new covariant form for energy-momentum transfer in dark sector}
\author{Cheng-Yi Sun\footnote{cysun@mailis.gucas.ac.cn; ddscy@163.com}\
$^{,1}$,\ Yu Song$^{1}$ and Rui-Hong Yue$^{2}$
\\
 {$^1$\small Institute of Modern Physics, Northwest University, Xian 710069, P.R.
 China.}\\
{$^2$\small Faculty of Science, Ningbo University, Ningbo 315211,
P.R. China.}}

\begin{document}
\maketitle
\begin{abstract}
It was found that the model with interaction between cold dark
matter (CDM) and dark energy (DE) proportional to the energy density
of CDM $\rho_m$ and constant equation of state of DE $w_d$ suffered
from instabilities of the density perturbations on the supper-Hubble
scales. Here we suggest a new covariant model for the
energy-momentum transfer between CDM and DE. Then using the
covariant model, we analyze the evolution of density perturbations
on the supper-Hubble scale. We find that the instabilities can be
avoided in the model with constant $w_d$ and interaction
proportional to $\rho_m$. Furthermore, we analyze the dominant
non-adiabatic mode in the radiation era and find that the mode grows
regularly.

\end{abstract}
\section{Introduction}
We are convinced by the increasing observations
\cite{Supernova,WMAP,LSS} that the present universe is dominated by
the so called dark energy (DE), which accounts for $\simeq70\%$ of
the critical mass density and has been pushing the universe into
accelerated expansion \cite{dark energy1,dark energy2}. And the
other main component in the universe is cold dark matter (CDM),
which accounts for $\simeq30\%$ of the critical mass density and
behaves as the pressureless dust. Then it is natural for us to
consider that the two dark components might interact mutually. And
furthermore it is found that an appropriate interaction can help to
alleviate the coincidence problem \cite{IDE0,IDE1}, namely why DE
and CDM are comparable in size exactly today \cite{dark energy2}.
Different interacting models of dark energy have been investigated
intensively \cite{IDE2,IDE3}.

Usually, in the literature, the model with interaction between DE
and CDM is described by the two continuity equations
\begin{align}
  \label{CLBgM}
  \dot{\rho}_m+3H\rho_m&=Q,\\
  \label{CLBgD}
  \dot{\rho}_d+3H(1+w_d)\rho_d&=-Q,
\end{align}
where $Q$ denotes the phenomenological interaction term between DE
and CDM; $\rho_m$ and $\rho_d$ are the energy densities of CDM and
DE respectively; $w_d\equiv p_d/\rho_d$ is the equation of state
parameter of DE; $p_d$ is the pressure density of DE;
$H\equiv\dot{a}/a$ is the Hubble parameter; $a(t)$ is the scale
factor in the Friedmann-Robertson-Walker (FRW) metric; a dot denotes
the derivative with respect to the cosmic time $t$. In the note we
do not allow the phantom case $w_d<-1$. Owing to the lack of the
knowledge of micro-origin of the interaction, usually the
interaction term is parameterized in a simple form as \cite{IDE0}
\begin{equation}
  \label{Qgeneral}
  Q=3H(\alpha\rho_m+\beta\rho_d),
\end{equation}
where $\alpha$ and $\beta$ are positive constants. The interaction
term $Q$ would influence not only the background dynamics of the
universe, but also the growth of the perturbations of the
cosmological fluids.

Recently, in Ref.\cite{0804.0232}, by modeling DE as a fluid with
constant $w_d>-1$, the authors investigated the evolution of the
linear density perturbations and have shown that the combination of
constant $w_d$ and the simple interaction form $Q$ given in
Eq.(\ref{Qgeneral}) leads to an instability: the curvature
perturbation on the super-Hubble scales blows up in the early
universe \cite{0804.0232}. The explicit models investigated in
Ref.\cite{0804.0232} included the two cases of $\beta=0$ and
$\alpha=\beta$ in Eq.(\ref{Qgeneral}). Further more, in
\cite{0901.3272}, it is concluded that the perturbations in the dark
energy become unstable for any model with constant $w_d>-1$ and
non-zero $\alpha$, no matter how small the parameter $\alpha$ is
made. In \cite{0807.3471,0901.3272,0901.1611}, the case of
$\alpha=0$ was surveyed and it was found that the instability can be
avoided if $\beta$ is made small enough. In \cite{0808.1646}, by
modeling DE as a quintessence field, the author found that the
instability can also be avoided even for the interaction $Q$
proportional to $\rho_m$.

Then it seems that the model with constant $w_d$ and non-zero
$\alpha$ in Eq.(\ref{Qgeneral}) is ruled out as a viable interacting
model. However, in the note we try to show that the evolution of the
density perturbations becomes regular even in the case of $\beta=0$
in Eq.(\ref{Qgeneral}) if we adopt a new covariant form for
energy-momentum transfer between DE and CDM.

The remain part of the note is organized as follows. Firstly, we
display our new covariant model for interaction in the dark sector.
Secondly, by assuming the universe filled only by DE and CDM, we
investigate the evolution of the density perturbations and show that
instability can be avoided. Thirdly, by considering the effects of
the radiation (photons and neutrinos) and baryons, we survey the
dominant non-adiabatic mode in the radiation era and show that no
non-adiabatic mode blows up. Finally, conclusions and discussions
are given.

\section{A Covariant Model For Dark-Sector Interaction}

The conservation laws (\ref{CLBgM}) and (\ref{CLBgD}) work well in
describing the background evolution of the universe. But in order to
study the evolution of the density perturbation, we need a covariant
form for the energy-momentum transfer between DE and CDM which holds
in an inhomogeneous universe and reduces to Eqs.(\ref{CLBgM}) and
(\ref{CLBgD}) in a homogeneous FRW universe. Usually, the covariant
form for energy-momentum transfer is taken to be
\cite{oldCovForm,0804.0232}
\begin{equation}
  \label{EMTAQ}
  \nabla_\nu T^{\mu\nu}_A=Q^\mu_A,\quad \sum_AQ^\mu_A=0,
\end{equation}
where $A=m,d$ to denote CDM and DE respectively. In
Ref.\cite{0804.0232}, it is assumed that
\begin{equation}
  \label{QMQD}
  Q^{\mu}_m=-Q^{\mu}_d=Qu^{\mu}_m,
\end{equation}
where $u^\mu_m$ is the four velocity of CDM. The conclusions in
\cite{0804.0232} are based on the above ansatz.

However, we find that there may exist the other natural covariant
form for the energy-momentum transfer. Let us show it. In the note,
we take the interaction $Q$ to be
\begin{equation}
  \label{Qalpha}
  Q=3\alpha H\rho_m,
\end{equation}
This is just the case of $\beta=0$ in Eq.(\ref{Qgeneral}). Then we
can rewrite Eqs.(\ref{CLBgM}) and (\ref{CLBgD}) as
\begin{align}
  \label{CLBgCDMEff}
  \dot{\rho}_m+3H(\rho_m+p^{\text{eff}}_m)&=0\\
  \label{CLBgDEEff}
  \dot{\rho}_d+3H(\rho_d+p^{\text{eff}}_d)&=0,
\end{align}
where
\begin{align}
  \label{pmEff}
  p^{\text{eff}}_m&\equiv-\alpha\rho_m,\\
  \label{pdEff}
  p^{\text{eff}}_d&\equiv p_d+\alpha\rho_m.
\end{align}
Motivated by these equations, we may define the effective
energy-momentum tensors  of CDM and DE respectively as
\begin{align}
  \label{EMTensorEffCDM}
  T_{\text{em}}^{\mu\nu}&=\rho_mu_m^\mu u_m^\nu+p^{\text{eff}}_m(u_m^\mu u_m^\nu+g^{\mu\nu}),\\
  \label{EMTensorEffDE}
  T_{\text{ed}}^{\mu\nu}&=\rho_du_d^\mu u_d^\nu+p^{\text{eff}}_d(u_d^\mu u_d^\nu+g^{\mu\nu}),
\end{align}
where $u_m^\mu$ and $u_d^\mu$ are the four velocities of CDM and DE
respectively. And the two effective energy-momentum tensors are
conserved respectively
\begin{equation}
  \label{CLEff}
  {T_{\text{em}}^{\mu\nu}}_{;\mu}={T_{\text{ed}}^{\mu\nu}}_{;\mu}=0
\end{equation}
It can be easily checked that Eqs.(\ref{CLBgM}) and (\ref{CLBgD})
can be reduced from Eq.(\ref{CLEff}) in the background FRW universe.
The corresponding Einstein equations should be
\begin{equation}
  \label{EinsteinEq}
  G^{\mu\nu}=8\pi G(T_{\text{em}}^{\mu\nu}+T_{\text{ed}}^{\mu\nu}).
\end{equation}
Now we can survey the evolution of the curvature perturbation by
expanding Eqs.(\ref{CLEff}) and (\ref{EinsteinEq}) to the first
order of the density perturbations.

The covariant model given in the last paragraph is very different
from the one defined in Eq.(\ref{EMTAQ}), although both of them give
the same background evolution of the universe. In the model in
Eq.(\ref{EMTAQ}), the perturbations of the interaction effect the
evolution of the density perturbations via the continuity equations
and do not appear in the Einstein equations explicitly, while in the
model defined in the last paragraph, the perturbations of the
interaction appears explicitly both in the continuity equations and
in the Einstein equations.

\section{Evolution of Density Perturbations}

In the section, we apply the covariant model defined in
Eqs.(\ref{CLEff}) and (\ref{EinsteinEq}) to study the evolution of
the density perturbations in the model with $Q$ given in
Eq.(\ref{Qalpha}) and constant $w_d$. For simplicity, we consider a
flat FRW universe filled only by DE and CDM. The perturbed FRW
metric in the conformal Newtonian gauge is given by
\begin{equation}
  \label{FRWFirOrd}
  ds^2=a^2(\tau)[-(1+2\phi)d\tau^2+(1-2\psi)d\textbf{x}^2],
\end{equation}
where $\phi$ and $\psi$ denote the scalar perturbations. The
corresponding Friedmann equation can be rewritten as
\begin{equation}
  \label{FE}
  \mathcal{H}^2\equiv\Big(\frac{a'}{a}\Big)^2=\frac{8\pi
  G}{3}(\rho_m+\rho_d)a^2.
\end{equation}
Hereafter, primes denote the derivatives with respect to the
conformal time $\tau$. With Eq.(\ref{Qalpha}) and constant $w_d$,
the background continuity equations (\ref{CLBgM}) and (\ref{CLBgD})
can be solved exactly:
\begin{align}
  \label{CDMBg}
  \rho_m&=\rho_{m0}a^{-3(1-\alpha)},\\
  \label{DEBg}
  \rho_d&=\rho_{d0}a^{-3(1+w_d)}+\Big(\frac{\alpha}{\alpha+w_d}\Big)\rho_{m0}a^{-3}(a^{-3w_d}-a^{3\alpha}).
\end{align}
Hereafter, the subscript $0$ denotes the present value of the
corresponding parameter and $a_0=1$.

\subsection{Evolving Equations of Density Perturbations}

When the perturbed metric in Eq.(\ref{FRWFirOrd}) is considered, the
four velocities of CDM and DE are
\begin{equation}
  \label{4VelocityPert}
  u^\mu_m=a^{-1}\Big(1-\phi,\partial_iv_m\Big),\quad
  u^\mu_d=a^{-1}\Big(1-\phi,\partial_iv_d\Big),
\end{equation}
where $v_m$ and $v_d$ are the peculiar velocity potentials of CDM
and DE respectively. Usually, we define the volume expansion rates
of CDM and DE (in Fourier space) respectively as
\begin{equation}
  \label{theta}
  \theta_m=-k^2v_m,\quad \theta_d=-k^2v_d.
\end{equation}
We use $\delta\rho_m$, $\delta\rho_d$, $\delta p_d$, $\delta
p^{\text{eff}}_m$ and $\delta p^{\text{eff}}_d$ to denote the
first-order perturbations of the corresponding parameters. Then we
can introduce the dimensionless factional density perturbations of
CDM and DE as
\begin{equation}
  \label{delta}
  \delta_m=\frac{\delta\rho_m}{\rho_m},\quad
  \delta_d=\frac{\delta\rho_d}{\rho_d}.
\end{equation}
The curvature perturbation on the constant-$\rho_A$ ($A=m,d$)
surface and the total curvature perturbation on the constant-$\rho$
($\rho=\rho_m+\rho_d$) surface are defined respectively as
\begin{equation}
  \label{curvPertCDMDE}
  \zeta_A=-\psi-\mathcal{H}\frac{\delta\rho_A}{\rho'_A},\quad
  \zeta=-\psi-\mathcal{H}\frac{\delta\rho_m+\delta\rho_d}{\rho'_m+\rho'_d}.
\end{equation}

From Eq.(\ref{pmEff}), we have
\begin{equation}
  \label{deltaPmEff}
  \delta p^{\text{eff}}_m=-\alpha\rho_m\delta_m.
\end{equation}
And from Eq.(\ref{pdEff}), we have
\begin{equation}
  \label{deltaPdEff}
  \delta p^{\text{eff}}_d=\delta p_d+\alpha\rho_m\delta_m.
\end{equation}
Here, following the analysis in Ref.\cite{0804.0232}, we take
\begin{equation}
  \label{deltaPd}
  \delta
  p_d=\delta\rho_d+(1-w_d)[3\mathcal{H}(1+w_d+\alpha\frac{\rho_m}{\rho_d})\rho_d]\frac{\theta_d}{k^2}.
\end{equation}
Then from Eq.(\ref{CLEff}), we can get the evolving equations of the
density and velocity perturbations of CDM and DE as
\begin{align}
  \label{dDeltaMdtau}
  \delta'_m+(1-\alpha)\theta_m-3(1-\alpha)\psi'&=0\\
  \label{dThetaMdtau}
  \theta_m'+\mathcal{H}(1+3\alpha)\theta_m-k^2\phi+k^2\frac{\alpha}{1-\alpha}\delta_m&=0
\end{align}
and
\begin{align}
  \label{dDeltaDdtau}
  \begin{split}
    \delta'_d+3\mathcal{H}(1-w_d-\alpha\frac{\rho_m}{\rho_d})\delta_d
             +9\mathcal{H}^2(1-w_d)(1+w_d+\alpha\frac{\rho_m}{\rho_d})\frac{\theta_d}{k^2}\quad\quad\quad\quad\quad\quad\quad\quad\quad&\\
             +3\alpha\mathcal{H}\frac{\rho_m}{\rho_d}\delta_m+(1+w_d+\alpha\frac{\rho_m}{\rho_d})(\theta_d-3\psi')&=0
  \end{split}\\
  \label{dThetaDdtau}
  \theta_d'-2\mathcal{H}\theta_d-3\mathcal{H}\frac{\alpha(1-\alpha)\frac{\rho_m}{\rho_d}}{1+w_d+\alpha\frac{\rho_m}{\rho_d}}\theta_d
             -k^2\phi-k^2\frac{\alpha\frac{\rho_m}{\rho_d}}{1+w_d+\alpha\frac{\rho_m}{\rho_d}}\delta_m
             -k^2\frac{\delta_d}{1+w_d+\alpha\frac{\rho_m}{\rho_d}}&=0
\end{align}

In the conformal Newtonian gauge, the first-order perturbed Einstein
equations Eq.(\ref{EinsteinEq}) give us \cite{9506072}
\begin{align}
  \label{EinsteinEq00}
  &3\mathcal{H}\psi'+k^2\psi+3\mathcal{H}^2\phi=-4\pi Ga^2(\delta_m\rho_m+\delta_d\rho_d),\\
  \label{EinsteinEq0i}
  &k^2\psi'+k^2\mathcal{H}\phi=4\pi Ga^2[(1-\alpha)\rho_m\theta_m+(1+w_d+\alpha\frac{\rho_m}{\rho_d})\rho_d\theta_d]\\
  \label{EinsteinEqii}
  &\psi''+\mathcal{H}(2\psi'+\phi')+(2\frac{a''}{a}-\mathcal{H}^2)\phi+\frac{k^2}{3}(\psi-\phi)=4\pi
                   Ga^2(\delta p^{\text{eff}}_m+\delta p^{\text{eff}}_d)\\
  \label{EinsteinEqij}
  &\psi-\phi=0.
\end{align}
Only two of the above equations are independent. Choosing two of
them (e.g. Eqs.(\ref{EinsteinEq00}) and (\ref{EinsteinEqij})), and
using Eqs.(\ref{dDeltaMdtau}), (\ref{dThetaMdtau}),
(\ref{dDeltaDdtau}) and (\ref{dThetaDdtau}) together, we can solve
these evolving equations numerically, if the initial conditions are
given.

\subsection{Adiabatic Initial Conditions}

In the early universe, $a\ll1$, Eqs.(\ref{CDMBg}) and (\ref{DEBg})
indicate
\begin{equation}
  \label{rhoMOverRhoD}
  \frac{\rho_m}{\rho_d}\rightarrow-\frac{w_d+\alpha}{\alpha},
\end{equation}
and then, from Eq.(\ref{FE}), we have
\begin{equation}
  \label{initialMDHTau}
  \mathcal{H}=\frac{2}{1-3\alpha}\tau^{-1},\quad
  \tau=\frac{2}{(1-3\alpha)H_0}\sqrt{\frac{w_d+\alpha}{w_d\Omega_{m0}}}a^{\frac{1}{2}(1-3\alpha)},
\end{equation}
where
\[
  \Omega_{m0}\equiv\frac{8\pi G\rho_{m0}}{3H_0^2}.
\]
Here we adopt the adiabatic initial conditions to study the
evolution of the density perturbations on the super-Hubble scales
($k\ll aH$). To the lowest order in $k\tau$, we can set the
adiabatic conditions as
\begin{align}
  \label{initialPhiDis}
  \phi&=\psi=A_\phi=\text{Const.},\\
  \label{initialDelta}
  \delta_m&=\delta_d=-\frac{2w_d}{w_d+\alpha}A_\phi,\\
  \label{initialThetaM}
  \theta_m&=\frac{(w_d+\alpha+\alpha w_d-\alpha^2)(1-3\alpha)}{3(1-\alpha^2)(w_d+\alpha)}k^2\tau A_\phi,\\
  \label{initialThetaD}
  \theta_d&=\frac{[2w_d^2-(1-\alpha)(w_d-\alpha)](1-3\alpha)}{3(1-\alpha)(w_d+\alpha)(2w_d+\alpha-1)}k^2\tau A_\phi.
\end{align}

\subsection{Evolution of Curvature Perturbation}

\begin{figure}
\centering
\renewcommand{\figurename}{Fig.}
%\captionstyle{small}
\includegraphics[scale=0.8]{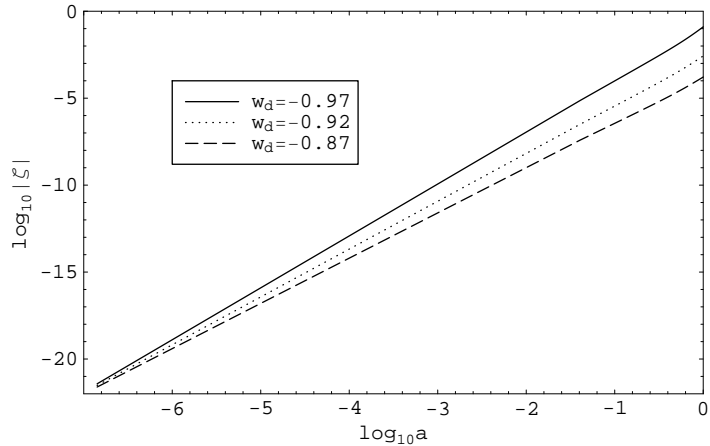}
\caption{$\log_{10}{|\zeta|}$ versus $\log_{10}{a}$ in the
interacting model for fixed $\Omega_{m0}=0.3$, $h=0.67$,
$k=1.5\times10^{-4}\text{Mpc}^{-1}$, $\alpha=10^{-3}$ and different
$w_d$.\label{FigZetaWd}}
\end{figure}

\begin{figure}
\centering
\renewcommand{\figurename}{Fig.}
%\captionstyle{small}
\includegraphics[scale=0.8]{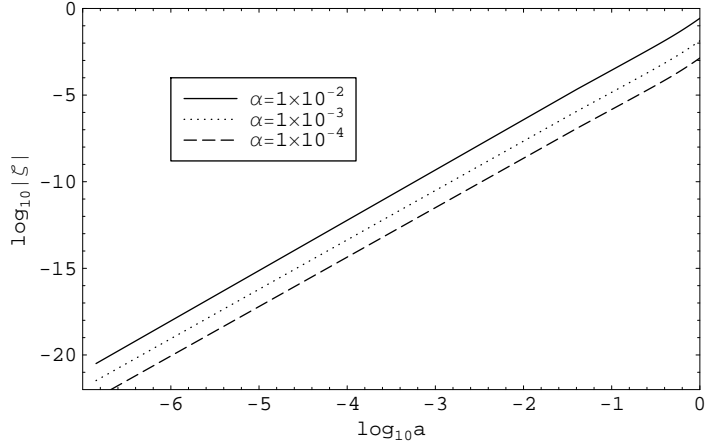}
\caption{$\log_{10}{|\zeta|}$ versus $\log_{10}{a}$ in the
interacting model for fixed $\Omega_{m0}=0.3$, $h=0.67$,
$k=1.5\times10^{-4}\text{Mpc}^{-1}$, $w_d=-0.94$ and different
$\alpha$.\label{FigZetaAlpha}}
\end{figure}

Now using the initial conditions given in the last subsection, we
can solve the evolving equations
(\ref{dDeltaMdtau})-(\ref{EinsteinEq00}) and (\ref{EinsteinEqij})
numerically to obtain the evolution of the density perturbation and
then get the evolution of the curvature perturbation $\zeta$ by
using Eq.(\ref{curvPertCDMDE}). We have displayed the results in
Fig.\ref{FigZetaWd} and Fig.\ref{FigZetaAlpha}. In the two figures,
we have fixed the parameters as $\Omega_{m0}=0.3$,
$k=1.5\times10^{-4}\text{Mpc}^{-1}$,
$H_0=100h\;\text{km}\;\text{sec}^{-1}\text{Mpc}^{-1}$ and $h=0.67$.
And we have taken $A_\phi=10^{-25}$ and the initial values of $a$ to
be $a=10^{-11}$. In Fig.\ref{FigZetaWd}, we show the evolutions of
$\log_{10}|\zeta|$ for fixed $\alpha=1\times10^{-3}$ and different
$w_d$. And in Fig.\ref{FigZetaAlpha} we show the evolutions of
$\log_{10}|\zeta|$ for fixed $w_d=-0.94$ and different $\alpha$. The
evolutions displayed in Fig.\ref{FigZetaWd} and
Fig.\ref{FigZetaAlpha} manifest the standard power-law growth and no
instabilities are present, which are similar the results in
\cite{0808.1646}.

\section{Dominant Non-Adiabatic Mode}

In the last section, we have shown that in our new covariant model,
instabilities on the super-Hubble scale can be avoided. But the
conclusion is obtained by assuming that the universe is filled only
by CDM and DE. In this section, we discuss the dominant
non-adiabatic mode deep in the radiation-dominated era by including
the components of photons, neutrinos and baryons in the universe. If
the dominant non-adiabatic mode evolves regular in the radiation
era, we believe the instabilities can also be avoided even when the
components of radiations and baryons are involved.

The $A$-fluid energy-momentum tensor including perturbations is
taken to be
\begin{equation}
  \label{EMTA}
  T^{\mu\nu}_A=(\rho_A+p_A)u^{\mu}_Au^{\nu}_A+p_Ag^{\mu\nu}+\pi^{\mu\nu}_A,
\end{equation}
where $u^\mu_A$ is the four velocity
\[
  u^{\mu}_A=a^{-1}\Big(1-\phi,\partial_iv_A\Big).
\]
We have allowed an anisotropic shear perturbation $\pi^{\mu\nu}_A$,
and $A=b,\gamma,\nu$ to denote the corresponding parameters of
baryons, photons and neutrinos. We take
$\pi_b^{\mu\nu}=\pi_\gamma^{\mu\nu}=0$ and
\begin{equation}
  \label{piNeutrino}
  \pi_\nu^{0\mu}=0,\quad
  \pi^{ij}_\nu=a^{-2}\Big(\partial_i\partial_j-\frac{1}{3}\delta^{ij}\big)\pi_\nu.
\end{equation}
The fluids of CDM and DE are described by the effective
energy-momentum tensors defined in Eqs.(\ref{EMTensorEffCDM}) and
(\ref{EMTensorEffDE}) respectively.

Early in the radiation era, the Friedmann equation reads
\begin{equation}
  \label{FERD}
  \mathcal{H}^2a^{-2}=\frac{8\pi G}{3}(\rho_\gamma+\rho_\nu)=\frac{8\pi
  G}{3}\rho_{r0}a^{-4}.
\end{equation}
Then we have
\begin{equation}
  \label{initialRDHTau}
  a=\sqrt{\Omega_{r0}}H_0\tau,\quad \mathcal{H}=\tau^{-1}, \quad
  \Omega_{r0}\equiv\frac{8\pi G\rho_{r0}}{3H^2_0}.
\end{equation}
In the radiation era, the perturbed Einstein equations give us that
\begin{align}
  \label{EinsteinEqRD00}
  &3\tau^{-1}\psi'+k^2\psi+3\tau^{-2}\phi=-4\pi Ga^2\Big(\delta\rho_m+\delta\rho_d+\sum_A\delta\rho_A\Big),\\
  \label{EinsteinEqRD0i}
  &k^2(\psi'+\tau^{-1}\phi)=4\pi Ga^2\Big[(\rho_m+p^{\text{eff}}_m)\theta_A+(\rho_d+p^{\text{eff}}_d)\theta_A+\sum_A(\rho_A+p_A)\theta_A\Big],\\
  \label{EinsteinEqRDii}
  &\psi''+2\tau^{-1}\psi'+\tau^{-1}\phi'-\tau^{-2}\phi+\frac{k^2}{3}(\psi-\phi)=4\pi
    Ga^2\Big(\delta
    p^{\text{eff}}_m+\delta p^{\text{eff}}_d+\sum_A\delta p_A\Big),\\
  \label{EinsteinEqRDij}
  &\psi-\phi=8\pi G\pi_\nu,
\end{align}
where $A$ runs over $b,\gamma,$ and $\nu$.

For CDM and DE, the perturbed continuity equations are given by
Eqs.(\ref{dDeltaMdtau})-(\ref{dThetaDdtau}). For baryons, the
perturbed continuity equations (in Fourier space) are
\cite{0804.0232}
\begin{align}
  \label{dDeltaBdtau}
  \delta_b'&=-\theta_b+3\psi',\\
  \label{dThetaBdtau}
  \theta_b'&=-\mathcal{H}\theta_b+k^2\phi,
\end{align}
and for photons \cite{0804.0232}
\begin{align}
  \label{dDeltaGammadtau}
  \delta_\gamma'&=-\frac{4}{3}\theta_\gamma+4\psi',\\
  \label{dThetaGammadtau}
  \theta_\gamma'&=\frac{1}{4}k^2\delta_\gamma+k^2\phi,
\end{align}
and for neutrinos \cite{0804.0232}
\begin{align}
  \label{dDeltaNudtau}
  \delta_\nu'&=-\frac{4}{3}\theta_\nu+4\psi',\\
  \label{dThetaNudtau}
  \theta_\nu'&=\frac{1}{4}k^2\delta_\nu+k^2\phi-k^2\sigma_\nu,\\
  \label{dSigmaNudtau}
  \sigma_\nu'&=\frac{4}{15}\theta_\nu,
\end{align}
where $\theta_A=-k^2v_A$ and
$\sigma_\nu\equiv2k^2\pi_\nu/[3a^2(\rho_\nu+p_\nu)]$

In order to find the dominant non-adiabatic mode, we assume a
leading-order power law for perturbations \cite{0804.0232}
\begin{equation}
  \label{defineNonAdiaMod}
  \psi=A_\psi(k\tau)^{n_\psi},\ \phi=A_\phi(k\tau)^{n_\phi},\
  \delta_A=B_A(k\tau)^{n_A},\ \theta_A=C_A(k\tau)^{s_A},\
  \sigma_\nu=D_\nu(k\tau)^{n_\sigma}.
\end{equation}
Here the subscript $A=c,d,b,\gamma$, and $\nu$ denotes the
corresponding parameter of CDM, DE, baryons, photons and neutrinos
respectively. To the leading order in $k\tau$, the equations
(\ref{dDeltaMdtau})-(\ref{dThetaDdtau}) and
(\ref{EinsteinEqRD00})-(\ref{dSigmaNudtau}) may be solved, in terms
of $\psi$:
\begin{align}
  \label{DisplayPhiPsi}
  \phi&=J\psi,\quad J=1-\frac{16R_\nu}{5(n_\psi+2)(n_\psi+1)+8R_\nu},\\
  \label{DisplayDeltaThetaGammaPsi}
  \delta_\gamma&=\delta_\nu=4\psi,\quad
  \theta_\gamma=\theta_\nu=\frac{J+1}{n_\psi+1}k^2\tau\psi,\\
  \label{DisplayDeltaThetaBPsi}
  \delta_b&=3\psi,\quad
  \theta_b=\frac{J}{n_\psi+2}k^2\tau\psi,\\
  \label{DisplayDeltaThetaCDMPsi}
  \delta_m&=3(1-\alpha)\psi,\quad
  \theta_m=\frac{J-3\alpha}{n_\psi+2+3\alpha}k^2\tau\psi,\\
  \label{DisplayDeltaDPsi}
  \delta_d&=\frac{2\Omega_{r0}^{(1-3\alpha)/2}}{\alpha\Omega_{m0}H_0^{1+3\alpha}}(w_d+\alpha)(n_\psi+J+2)\frac{\psi}{\tau^{1+3\alpha}},\\
  \label{DisplayThetaDPsi}
  \theta_d&=-\frac{n_\psi+2}{9(1-w_d)(1-\alpha)}k^2\tau\delta_d,
\end{align}
where $R_\nu\equiv\rho_\nu/(\rho_\gamma+\rho_\nu)$ and
\begin{equation}
  \label{DisplayNpsi}
  n_\psi=\frac{-3w_d\pm\sqrt{9w_d^2+12w_d-20}}{2},
\end{equation}
Eq.(\ref{DisplayDeltaDPsi}) indicates that the modes are regular
(i.e. well behaved as $k\tau\rightarrow0$) provided
\[
  \text{Re}[n_\psi]\ge1+3\alpha.
\]
For $w_d\sim-1$, this leads to
\[
  -\frac{3}{2}w_d\ge1+3\alpha\Rightarrow\alpha\lesssim\frac{1}{6}.
\]

Correspondingly, the total curvature perturbation $\zeta$ is defined
as
\[
  \zeta=-\psi-\mathcal{H}\frac{\sum_A\delta\rho_A}{\sum_A\rho_A'}
\]
where $A$ runs over $m,d,\gamma,\nu$ and $b$. Then $\zeta$ can be
expressed in terms of $\psi$ as
\begin{equation}
  \label{zetaRDLeadingOrder}
  \zeta=-\frac{1}{2}(n_\psi+J+2)\psi.
\end{equation}
For $w_d\sim-1$, $n_\psi$ is a complex number and
\[
  \text{Re}[n_\psi]\sim\frac{3}{2}.
\]
So the dominant non-adiabatic mode grows in a regular power law and
no instabilities are present.

\section{Conclusions and Discussions}

In the note, we have suggested a new covariant model for dark-sector
interaction to avoid the instabilities of the curvature perturbation
on the supper-Hubble scale. By using the covariant model and
choosing $Q=3\alpha H\rho_m$ and constant $w_d$, we analyze the
evolution of density perturbations in a universe filled only by CDM
and DE and find that the instabilities shown in \cite{0804.0232} can
be avoided in our covariant model. Further more, we analyze the
dominant non-adiabatic mode early in the radiation era, and find the
non-adiabatic mode evolves regularly. So we believe that in our
covariant model, the instabilities can also be avoided even in the
universe filled by radiation, matter and DE.

Actually, it is not very surprising for the instabilities to be
avoided in our covariant model. We know, from
Eqs.(\ref{EMTensorEffCDM})-(\ref{EinsteinEq}), that the two
mutually-interacting dark fluids can be described respectively by
two effective energy-momentum tensors which are conserved
separately. This means the interacting model can be taken as a
non-interacting model effectively, while no instabilities are
present in non-interacting models. So we expect the instabilities
can be avoided in our covariant model.

\section*{Acknowledgments}
This work has been supported in part by the Research Fund for the
Doctoral Program of Higher Education of China under Grant No.
20106101120023, the National Natural Science Foundation of China
under Grant No. 10875060, and the Natural Science Foundation of the
Northwest University of China under Grant No. 09NW27.


\begin{thebibliography}{99}
\bibitem{Supernova}
A. G. Riess \emph{et al.}, Astron. J. 116, 1009 (1998),
[astro-ph/9805201]; S. Perlmutter \emph{et al.}, Astrophys. J. 517,
565 (1999), [astro-ph/9812133].

\bibitem{WMAP}
D. N. Spergel \emph{et al.}, Astrophys. J. Suppl. 148, 175 (2003),
[astro-ph/0302209]; D. N. Spergel \emph{et al.}, Astrophys. J.
Suppl. 170, 377 (2007), [astro-ph/0603449].

\bibitem{LSS}
M. Tegmark \emph{et al.}, Phys. Rev. D 69, 103501 (2004),
[astro-ph/0310723]; K. Abazajian \emph{et al.}, Astron. J. 128, 502
(2004), [astro-ph/0403325]; K. Abazajian \emph{et al.}, Astron. J.
129, 1755 (2005), [astro-ph/0410239].

\bibitem{dark energy1}
S. Weinberg, Rev. Mod. Phys. 61, 1 (1989);

\bibitem{dark energy2}
V. Sahni and A. A. Starobinsky, Int. J. Mod. Phys. D 9, 373 (2000),
[astro-ph/9904398]; S. M. Carroll, Living Rev. Rel. 4, 1 (2001),
[astro-ph/0004075]; P. J. E. Peebles and B. Ratra, Rev. Mod. Phys.
75, 559 (2003), [astro-ph/0207347]; T. Padmanabhan, Phys. Rept. 380,
235 (2003), [hep-th/0212290]; E. J. Copeland, M. Sami and S.
Tsujikawa, Int. J. Mod. Phys. D 15, 1753 (2006), [hep-th/0603057];
R. Bousso, Gen. Rel. Grav. 40, 607 (2008), arXiv:0708.4231[hep-th].

\bibitem{IDE0} L. P. Chimento, Journal of Mathematical Physics 38, 2565
(1997), [arXiv:physics/9702029]; J. D. Barrow and T. Clifton, Phys.
Rev. D 73, 103520 (2006), [gr-qc/0604063].

\bibitem{IDE1}L. Amendola, Phys. Rev. D 62, 043511 (2000),
[arXiv:astro-ph/9908023]; L. P. Chimento, A. S. Jakubi, D. Pavon and
W. Zimdahl, Phys. Rev. D 67, 083513 (2003),
[arXiv:astro-ph/0303145]; R. G. Cai and A. Wang, JCAP 0503, 002
(2005), [arXiv:hep-th/0411025]; G. Olivares, F. Atrio-Barandela and
D. Pavon, Phys. Rev. D 74, 043521 (2006), [arXiv:astro-ph/0607604].

\bibitem{IDE2}L. P. Chimento, Phys. Rev. D 81, 043525 (2010), arXiv:0911.5687[astro-ph.CO];
J. H. He and B. Wang, JCAP 0806, 010 (2008),
arXiv:0801.4233[astro-ph]; M. Szydlowski, T. Stachowiak and R.
Wojtak, Phys. Rev. D 73, 063516 (2006), [arXiv:astro-ph/0511650]; L.
P. Chimento, M. Forte and G. M. Kremer, Gen. Rel. Grav. 41, 1125
(2009), arXiv:0711.2646[astro-ph]; R. Bean, E. E. Flanagan, I.
Laszlo and M. Trodden, Phys. Rev. D 78, 123514 (2008),
arXiv:0808.1105[astro-ph]; C. Feng, B.Wang, Y. Gong and R. K. Su,
JCAP 0709, 005 (2007), arXiv:0706.4033[astro-ph]; L. Zhang, J. Cui,
J. Zhang and X. Zhang, Int. J. Mod. Phys. D 19, 21 (2010),
arXiv:0911.2838[astro-ph.CO].

\bibitem{IDE3} S. Micheletti, E. Abdalla and B. Wang, Phys. Rev. D 79,
123506 (2009), arXiv:0902.0318[gr-qc]; S. Chen, B. Wang and J. Jing,
Phys. Rev. D 78, 123503 (2008), arXiv:0808.3482[gr-qc]; H. Wei and
R. G. Cai, Eur. Phys. J. C 59, 99 (2009), arXiv:0707.4052[hep-th];
H. Wei and R. G. Cai, Phys. Lett. B 655, 1 (2007),
arXiv:0707.4526[gr-qc]; M. Quartin, M. O. Calvao, S. E. Joras, R. R.
R. Reis and I. Waga, JCAP 0805, 007 (2008),
arXiv:0802.0546[astro-ph].

\bibitem{0804.0232}J. Valiviita, E. Majerotto, and R. Maartens,
%``Large-scale instability in interacting dark energy and dark matter fluids",
JCAP 0807, 020 (2008), arXiv:0804.0232[astro-ph].

\bibitem{0807.3471}J. H. He, B. Wang, and E. Abdalla,
%``Stability of the curvature perturbation in dark sectors¡¯ mutual interacting models",
Phys. Lett. B 671, 139 (2009), arXiv:0807.3471[gr-qc].

\bibitem{0901.3272}B. M Jackson and A. Taylor,
%``On the large-scale instability in interacting dark energy and dark matter fluids",
Phys. Rev. D 79, 043526 (2009), arXiv:0901.3272[astro-ph.CO].

\bibitem{0901.1611}M. B. Gavela, D. Hernandez, L. L. Honorez, O.
Mena, and S. Rigolin,
%``Dark coupling",
JCAP 0907, 034 (2009), arXiv:09011.611[astro-ph].

\bibitem{0808.1646}P. S. Corasaniti,
%``Slow-Roll Suppression of Adiabatic Instabilities in Coupled Scalar Field-Dark Matter",
Phys. Rev. D 78, 083538 (2008), arXiv:0808.1646[astro-ph].

\bibitem{oldCovForm}H. Kodama and M. Sasaki, Prog. Theor. Phys. Suppl. 78, 1
(1984); P. K. S. Dunsby, M. Bruni and G. F. R. Ellis, Astrophys. J.
395, 54 (1992).

\bibitem{9506072}C. P. Ma and E. Bertschinger,
%``Cosmological Perturbation Theory in the Synchronous and Conformal Newtonian Gauges",
[arXiv:astro-ph/9506072].

\end{thebibliography}
\end{document}